\documentclass[twocolumn]{pasj00}

\begin{document}
\SetRunningHead{Miyawaki et al.}{Suzaku Observations of M82 X-1}
\Received{2008/07dd}
\Accepted{2008/09/dd}

\title{Suzaku Observations of M82 X-1 : \\ Detection of a Curved Hard X-ray Spectrum}

%

%
 \author{%
   Ryohei \textsc{Miyawaki},\altaffilmark{1}
   Kazuo \textsc{Makishima},\altaffilmark{1,2,3}
   Shin'ya \textsc{Yamada},\altaffilmark{1}
   Poshak \textsc{Gandhi},\altaffilmark{2}\\
   Tsunefumi \textsc{Mizuno},\altaffilmark{4}
   Aya \textsc{Kubota},\altaffilmark{5}
   Takeshi \textsc{Tsuru},\altaffilmark{6}
      and
   Hironori \textsc{Matsumoto}\altaffilmark{6}
}

\altaffiltext{1}{
   Department of Physics, The University of Tokyo,
   7-3-1 Hongo, Bunkyo-ku, Tokyo 113-0033}
\altaffiltext{2}{
   Cosmic Radiation Laboratory, Institute of Physical and Chemical 
   Research (RIKEN),\\
   2-1 Hirosawa, Wako-shi, Saitama 351-0198}
\altaffiltext{3}{
   Research Center for the Early Universe (RESCUE), The University of Tokyo,\\
   7-3-1 Hongo, Bunkyo-ku, Tokyo 113-0033}
\altaffiltext{4}{
Department of Physical Science, School of Science, Hiroshima University,\\
1-3-1 Kagamiyama, Higashi-Hiroshima, Hiroshima 739-8526}
\altaffiltext{5}{
Department of Electronic Information Systems, Shibaura Institute of Technology,\\
307 Fukasaku, Minuma-ku, Saitama-shi, Saitama 337-8570}
\altaffiltext{6}{
   Department of Physics, Kyoto University, Kitashirakawa-Oiwake-cho, \\
   Sakyo-ku, Kyoto 606-8502}
\email{maxima@phys.s.u-tokyo.ac.jp}


\maketitle

\begin{abstract}
A report is presented on Suzaku  observations 
of the ultra-luminous X-ray source X-1 in the starburst galaxy M82,
made  three time in 2005 October for an exposure of $\sim 30$ ks each.
The XIS signals from a  region of radius $3'$ around the nucleus 
defined a 2--10 keV flux of $2.1 \times 10^{-11}$ erg s$^{-1}$ cm$^{-2}$
attributable to point sources.
The 3.2--10 keV spectrum was slightly more convex
than a power-law with a photon index of 1.7.
In all  observations,
the HXD also detected signals from M82 up to $\sim 20$ keV,
at a 12--20 keV flux of $4.4 \times 10^{-12}$ erg s$^{-1}$ cm$^{-2}$.
The HXD spectrum was  steeper than that of the XIS.
The XIS and HXD spectra can be jointly reproduced by a cutoff power-law model, 
or  similar curved models.
Of the detected wide-band signals, 1/3 to 2/3 are attributable to X-1,
while the remainder to other discrete sources in  M82.
Regardless of the modeling of these contaminants,
the spectrum attributable to X-1 is more curved than a  power-law,
with  a bolometric luminosity of $ (1.5-3) \times 10^{40}$ erg s$^{-1}$.
These results are interpreted as Comptonized emission 
from a black hole of $100-200$ solar masses,
radiating roughly at the Eddington luminosity.
\end{abstract}

\section{Introduction}
\label{sec:intro}

Ultra-luminous compact X-ray sources (ULXs; e.g., \cite{makishima2000}), 
found in many nearby galaxies, 
are often considered as candidate  intermediate-mass black holes (BHs),
because their typical luminosity reaching $3 \times 10^{39}$ to $10^{41}$  erg s$^{-1}$,
calculated assuming isotropic radiation,
largely exceeds the Eddington limit for known stellar-mass BHs.
Since the first discovery by the {\rm Einstein} satellite  \citep{fabbiano1989},
ULXs have been studied extensively in soft X-rays with 
{\rm ROSAT} \citep{colbert1999,colbert2002,liu2005, liu2006},
and then in energies up to  $\sim 10$ keV with 
{\rm ASCA} \citep{makishima2000,mizuno2001},
{\rm Chandra} (e.g. \cite{swartz2004}),
{\rm XMM-Newton} (e.g. \cite{feng+kaaret2005, stobbart2006}),
and {\rm Suzaku} \citep{mizuno2007,isobe2008}.
Searches for their counterparts have also been carried out 
in the optical (e.g. \cite{pakull2003, ptak2006}) and 
radio (e.g. \cite{sanchez-sutil2006}) wavelengths.
In contrast, we have little information on ULXs in hard X-rays above 10 keV,
where the established mass-accreting BHs, 
including both stellar-mass ones and active galactic nuclei (AGNs),
are known to emit a considerable fraction of their radiative luminosity.

Among a number of ULXs known so far,
so-called source X-1 in the starburst galaxy M82 is the most extreme object,
as the  luminosity  reached $10^{41}$ erg s$^{-1}$.
Although the host galaxy M82 has been observed repeatedly with the past X-ray observatories
due to its nearby location ($\sim 3.6$ Mpc; \cite{freedman2004}) and its active star formation,
the bright diffuse thermal X-rays from this galaxy hampered the detection of X-1
till the {\rm ASCA} era.
The presence of this object, X-1,  was first suspected by \citet{tsuru1997},
based on differences of fluxes measured with EXSOAT, Ginga, and ASCA.
Subsequently, the object was confirmed as a variable hard source 
through  multiple {\rm ASCA} observations, 
and interpreted at first as an active galactic nucleus 
veiled by the thermal plasma emission \citep{matsumoto1999,ptak1999}.
However, thanks to the superb angular resolution of  {\rm Chandra},
it was later revealed to be a non-AGN source located $\sim 170$ pc 
off  the dynamical center of the galaxy \citep{matsumoto2001,kaaret2001,tsuru2004}.

To explain  the highest luminosity
($1\times 10^{41}$ erg s$^{-1}$ in 3--20 keV; \cite{rephaeli2002})
recorded from M82 X-1,
a mass of at least 700 $M_{\odot}$ is necessary 
assuming isotropic radiation with a sub-Eddington luminosity.
Moreover, the 54 mHz X-ray quasi-periodic oscillation (QPO),
detected with {\rm XMM-Newton} and {\rm RXTE} 
\citep{strohmayer2003,fiorito2004,mucciarelli2006,dewangan2006,kaaret2006},
suggests that the source, presumably a BH, has a mass of 100--300 $M_{\odot}$,
assuming the QPO frequency to scale inversely with the BH mass.
These  results,
together with a recent  study of variability pattern \citep{casella2008},
make M82 X-1 the most promising candidate for an intermediate-mass BH.
It is also suggested to have a 62 day periodicity \citep{kaaret2006}.

Spectral studies of M82 X-1 have been conducted extensively.
With {\rm ASCA} and {\rm XMM-Newton}, however,
such attempts were limited to relatively narrow bands, e.g., 2--10 keV,
because of the surrounding bright thermal X-rays.
As a result, the observed spectra were reproduced by a variety of spectral models,
including a power-law (PL) of photon index $\Gamma = 1.7-2.5$
\citep{matsumoto1999,dewangan2006},
a thermal plasma emission model with a temperature 4--11 keV \citep{matsumoto1999},
an unsaturated Comptonization model \citep{fiorito2004,agrawal2006},
and  slim-disk emission \citep{okajima2006}.
Since these modelings imply vastly different physical conditions of the emission region, 
the nature of M82 X-1 has remained ambiguous.
For example, the slim-disk modeling by \citet{okajima2006}
leads to a BH mass of $\sim 30~M_\odot$,
which is significantly lower than is implied by the other arguments.
Another complication with  X-1 is 
that it is heavily confused with other point sources in the core region of M82.

These confusion problems have been  resolved by {\rm Chandra}
with its high angular resolution.
In fact, \citet{kaaret2006} found 
that the 0.3--8 keV band  spectrum of X-1 can be 
represented by a PL model of a photon index $\Gamma \sim 1.7$.
Neverthelss, the information is still insufficient to arrive at a clear scenario
as to the nature of X-1.

Evidently, the detection of X-1 in harder energies is of crucial importance.
The Large Area Counter onboard {\rm Ginga}, 
sensitive over the 2--30 keV energy band,
detected signals up to $\sim$20 keV from the M82 galaxy,
and yielded a spectrum which is better described by a thermal bremsstrahlung model of
temperature $\sim 5.7$ keV  than by a single PL model \citep{tsuru1992}.
The emission,  interpreted at that time as arising from a thin hot plasma, 
is now more likely to be attributable primarily to X-1.
\citet{cappi1999}  successfully described a  3--30 keV BeppoSAX spectrum 
of  M82 by a thermal plasma model of temperature $\sim 8.2$ keV.
Furthermore, {\rm RXTE} observations  of M82 on 30 occasions
yielded 3--50 keV spectra \citep{rephaeli2002}; 
the spectra acquired in brighter and fainter phases were both reproduced by thermal plasma models, 
with a temperature  of $\sim 6.6$ and $\sim 7.4$ keV, respectively.
These existing results in harder energies commonly indicate 
mildly curving broad-band spectra, rather than a PL-like one.
However, the fields-of-view (FOVs) of these 
non-imaging instruments also contained the spiral
galaxy M81, hosting a low-luminosity AGN
\citep{ishisaki1996,pellegrini2000,laparola2004}, 
located $37'$ away from M82.
Therefore, possible contamination to the observed signal was not excluded.

The Hard X-ray Detector (HXD; \cite{takahashi2007,kokubun2007}) 
onboard {\rm Suzaku} \citep{mitsuda2007},
with its unprecedented sensitivity in a broad (10--600 keV) energy band
and a tightly collimated FOV of $34' \times 34'$ (FWHM),
is expected to provide improved hard X-ray information on M82 X-1.
Here, we report on the {\rm Suzaku} detection of M82 X-1 up to $\sim 20$ keV,
and interpret the results in terms of an intermediate-mass BH.


\section{Suzaku Observations and Data Reduction}
\label{sec:observations}

We observed M82 with {\rm Suzaku} three times 
in the Science Working Group (SWG) phase;
on 2005 October 4 UT11:50 through October 5 UT03:24, 
October 19 UT00:40--22:32, and October 27 UT11:05 through 28 UT02:24.
Aiming primarily at the study of hot-gas outflows from M82,
the three observations were all conducted with the XIS optical axis pointed at
$(\alpha^{2000}, \delta^{2000}) = (9^{\rm h}55^{\rm m}34^{\rm s}, 69^{\circ}45'53'')$,
which is $\sim 5'$ north of the galaxy.
The XIS results on this offset region have already been published as \citet{tsuru2007}.
In the present paper, we focus on the central region of this galaxy,
including in particular X-1.

\begin{figure}[htbp]
  \begin{center}
   \FigureFile(75mm,){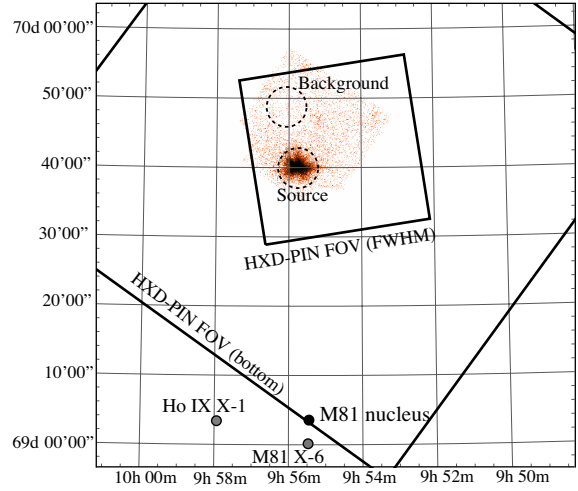}
\end{center}
  \caption{
    A background-inclusive 2--10 keV XIS~0  image from the second observation.
    The smaller and larger squares show the FWHM and full-width at
    zero transmission FOVs of HXD-PIN, respectively.
   Dashed circles show the source and background regions used for the XIS analysis.
    A filled small circle indicates the location of the M81 nucleus,
    whereas two gray circles those of the two ULXs in this field, 
    M81 X-6 and Holmberg IX X-1 \citep{fabbiano1988a}.
  }
  \label{fig:image}
\end{figure}

\begin{figure}[htbp]
  \begin{center}
    \FigureFile(160mm,){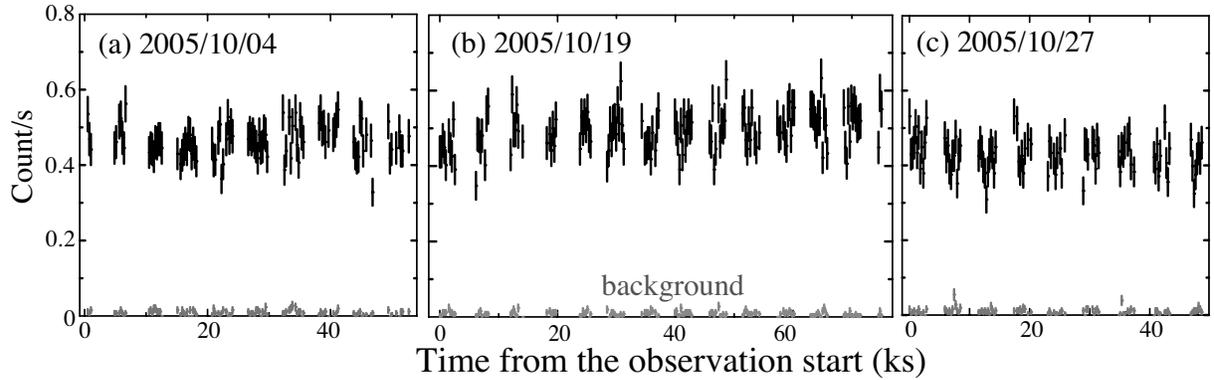}
  \end{center}
  \caption{
    XIS~0  light curves in the 2--10 keV band extracted from the source (black) and background (gray) 
    regions, acquired on (a) October 4, (b) October 19, and (c) October 27.
    The bin width is 256 s.
  }
  \label{fig:lc_xis}
\end{figure}

\begin{figure}[htbp]
\vspace*{-8mm}
  \begin{center}
    \FigureFile(75mm,){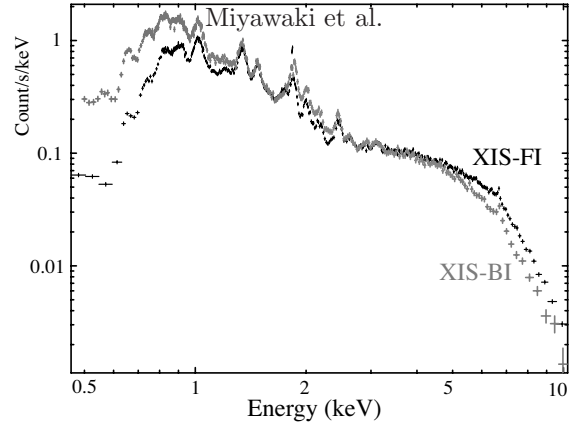}
 \end{center}
  \caption{Background-subtracted XIS spectra of the core region of M82,
   averaged over the three observations.
    The instrumental response is included.
    The FI (three cameras summed) and BI CCD data are shown in black and gray, respectively.}
  \label{fig:spec_xis_wide}
\end{figure}

Figure \ref{fig:image} shows an XIS~0 image in the 2--10 keV band 
from the second observation.
The X-ray bright part coincides with the central region of M82,
including luminous X-ray sources such as X-1.
The M81 nucleus during this observation was right on the edge of the HXD-PIN FOV, 
where the  transmission is $\leq 1\%$ 
even considering attitude uncertainties.
Similarly, the HXD-PIN transmission for M81 was $\leq 3\%$  
both in the first and third observations.
{As shown in the figure,
two ULXs known in the present sky region,
M81 X-6 and Holmberg XI X-1 (= M81 X-9; \cite{fabbiano1988a}),
were both outside the HXD-PIN FOV.

We used the software package HEADAS version 6.3.1 and
xselect version 2.4 to extract light curves, images, and spectra,
and xspec version 11 to analyze the spectra.
The XIS and HXD datasets were screened by the revision 1.2 
pipeline processing using the standard criteria as follows.
Events were discarded if they were acquired under the South Atlantic Anomaly (SAA),
or at regions of low cutoff rigidity 
($\leq 6$ GV for XIS and $\leq 8$ GeV c$^{-1}$ for HXD),
or with the Earth elevation angle of $< 5^{\circ}$, 
or during those periods when the telemetry was saturated.
The XIS events were further excluded 
when the source elevation above the sunlit Earth is $< 20^{\circ}$,
and finally events with grade 0, 2, 3, 4, and 6 were selected.
After these screenings, the three observations gave effective exposure of
29 ks, 36 ks, and 26 ks with the XIS, while 24 ks, 31 ks, and 22 ks with the HXD.


\section{Data Analysis and Results}

\subsection{XIS Data Analysis}
\label{subsec:XIS}

We extracted on-source XIS events from a region of $3'$ 
radius centered on the M82 nucleus,
and background events from another region of $3'$ radius
which is free from the diffuse X-ray emission associated with M82.
In figure \ref{fig:image}, these two regions are indicated by dashed circles.
We utilize events from all four CCD cameras,
after co-adding those from the three FI cameras (XIS~0, XIS~2, and XIS~3)
which have virtually identical responses.

\subsubsection{XIS light curves}
\label{subsubsec:XIS_ltcv}

Figure \ref{fig:lc_xis} shows 2--10 keV XIS~0 
light curves obtained in the three observations.
The background counts are negligible ($\leq 2\%$), 
compared to the signal counts which were determined as
$0.460 \pm 0.007$, $0.489 \pm 0.006$, and $0.425 \pm 0.007$ cts s$^{-1}$
in the three observations (the errors being 90\% confidence ranges).
Thus, the signal on the second and third occasions 
are higher by $\sim 6\%$ and lower by $\sim 8\%$,
respectively, than that  in the first observation.
By fitting these light curves  with a constant, we obtained 
$\chi^2/{\rm d.o.f}$ (degree of freedom) of 95.7/95, 190.0/131, and 110.1/87, respectively.
The probability of these signal counts being consistent 
with a constant value is $\sim$ 40\%, 1\%, and 5\%,
in the first, second, and third observations, respectively.
Therefore, the source is likely to have varied during the second observation;
in fact, the light curve suggests a gradual intensity increase by $\sim 10\%$
across the gross coverage of about one day.

\subsubsection{XIS spectra}

Figure \ref{fig:spec_xis_wide} shows 0.5--10 keV spectra 
from the FI CCD (black; averaged over XIS~0 , XIS~2, and XIS~3)
and the BI CCD (gray; XIS~1), obtained by accumulating events over the source region 
of figure \ref{fig:image}, and then subtracting those from the background region.
To grasp the rough spectral properties,
we tentatively summed up  the three observations.
Thus, the spectra are dominated by a rather hard continuum
at energies above $\sim 3$ keV,
whereas  by line-rich softer thermal emission at lower energies.
As confirmed through past observations,
the hard continuum must be coming from luminous point X-ray sources
including X-1 in particular,
while the soft thermal emission from the diffuse hot plasmas
(e.g., \cite{tsuru1997,moran1997,griffiths2000,stevens2003,ranalli2008,strickland2007}).
A rough evaluation using a thermal model and a PL continuum indicates
that the XIS spectra in energies above 3.2 keV 
(i.e., above  K$\alpha$ line at 3.14 keV from  H-like Ar)  are 
contaminated by the  thermal component  by no more  than 9\%. 
Since we are interested in X-1, 
below we  study the XIS spectra in the 3.2--10 keV range,
discarding their softer part.

In the hard-band XIS spectra, we observe an emission line structure at 6--7 keV;
this must be Fe-K  lines, as reported in previous observations
(e.g. \cite{tsuru1997,matsumoto1999,cappi1999, rephaeli2002,
strohmayer2003,agrawal2006}).
Using the {\rm Chandra} and {\rm XMM-Newton} data,
\citet{strickland2007} spectrally resolved the iron line complex
into 6.7 keV (from He-like iron) and 6.4 keV (from nearly neutral iron) components,
and showed that both are primarily distributed diffusely
rather than associated with discrete sources.

\subsection{HXD-PIN Data Analysis}
\label{subsec:HXD}

Through the method described in section 2,
we extracted the HXD-PIN data from the three observations.
Although the data covers an energy range of 10--70 keV,
we hereafter discard the lowest energy band of 10--12 keV,
where the data are often affected by increased noise.
Light curves and spectra are corrected for instrumental dead times
after \citet{takahashi2007} and \citet{kokubun2007}.
In the present paper,
we do not intend to analyze the HXD-GSO data,
since the signals from M82 are estimated to be below the GSO detection limit.
In addition, the HXD-GSO FOV  becomes progressively wider above $\sim 80$ keV,
and hence the contamination by M81 becomes difficult to eliminate.

\subsubsection{Background subtraction}
\label{subsubsec:PIN_background}

To perform background subtraction which is crucial for the HXD analysis,
we utilized non X-ray background (NXB) models provided 
by the HXD detector team \citep{kokubun2007}.
When the present data analysis was carried out \citep{miyawaki2008}, 
there were two NXB models available, 
so-called {\it bgd\_A} (Watanabe et al., suzakumemo-2007-01)
and {\it bgd\_D} (dead-time uncorrected; Fukazawa et al., suzakumemo-2007-02),
which employ different algorithms.
We produced fake background events using the two NXB models,
both under the same observing conditions as the actual data,
and both utilizing the same auxiliary information 
such as the measured upper-discriminator hit rates.
By processing the fake events through 
the same screening criteria as for the actual on-source data,
we derived two NXB  event sets  for each observation.
Then, {\it bgd\_A} and {\it bgd\_D} respectively gave 12--70 keV count rates of
0.46 and 0.45 cts s$^{-1}$ on October 4,
0.49 and 0.48 cts s$^{-1}$ on October 19, and
0.52 and 0.50 cts s$^{-1}$ on October 27.
Thus, the two NXB models agree within 4\%.
Hereafter, we employ {\it bgd\_A} for our spectral analysis,
because it  tends to systematically 
over-predict the NXB than the other model
in this particular observation,
and hence is expected to yield more conservative results
for sources like M82 X-1 that are close to the HXD-PIN detection limit.

We  examined the NXB modeling accuracy
utilizing Earth-occulted data,
assuming the Earth to be  ``dark" for HXD-PIN.
These data were extracted using the same criteria as for the on-source data,
except that the Earth elevation angle was required
to be $< -5^{\circ}$ instead of $> 5^{\circ}$.
The attained exposure was 5.3 ks on October 19 and 1.3 ks on October 27, 
while the October 4 data were free from Earth occultations.
Figure \ref{fig:spec_pin_earth} compares the observed
Earth-occultation data, with the synthetic {\it bgd\_A} spectra
which emulate the NXB to be observed during these occultation periods.
The observed and modeled NXB count rates in the 12--70 keV energy range are
$0.56 \pm 0.02$ and 0.55 cts s$^{-1}$ respectively on October 19, and 
$0.54 \pm 0.04$ and 0.54 cts s$^{-1}$ on October 27 likewise,
where the errors refer to 90\% confidence levels.
Statistical errors associated with the NXB model are neglected,
because the model produces 10 times larger number of fake events.
When the energy band  is limited to 12--20 keV,
where signals are detected as described later,
the observed and modeled NXB count rates become
$0.31 \pm 0.01$ and 0.31 cts s$^{-1}$ respectively on October 19, 
and  $0.30 \pm 0.03$ and 0.30 cts s$^{-1}$ on October 27 likewise.

As confirmed above, 
the Earth-occultation data and the modeled NXB data
agree well with each other within statistic errors,
which are  4\% and 10\%, on October 19 and 27, respectively.
This statement applies both to the 12--70 keV and 12--20 keV bands.
Further considering  the NXB modeling accuracy of HXD-PIN
which is generally $\sim \pm$5\% \citep{mizuno2007},
and the 4\% difference between the two NXB models,
we hereafter regard 
that the NXB model correctly reproduces the HXD-PIN NXB
within a systematic error of at most 4\%.

\subsubsection{HXD signal count rates}
\label{subsubsec:PIN_counts}

After the above preparations, 
we accumulated the on-source  HXD-PIN data of M82,
and derived 12--20 keV count rates from the three observations.
In table~\ref{tbl:cr_pin}, these results are compared
with  the NXB model predictions in the same energy band.
In all three observations,
the on-source data exhibit an excess count rate  by $\sim 0.04$ cts s$^{-1}$ 
above the modeled NXB.
Of these excess counts, 
we expect  0.018 cts s$^{-1}$ to be  coming from the cosmic X-ray background (CXB),
as calculated from the {\rm HEAO-1} results \citep{boldt1987,gruber1999}.
Further subtracting this CXB contribution,
we are finally left with  0.016--0.027 cts s$^{-1}$  (table~\ref{tbl:cr_pin})
in 12--20 keV to be attributable to the net signals from M82.
Although the 12--20 keV CXB intensity within the HXD-PIN FOV  
is expected to fluctuate by $\sim \pm 12\%$ (rms) from sky to sky,
this is  only to $\pm  1\%$ of the NXB in the same energy range,
because the average CXB count rate therein, 
typically $7\%$ of the NXB, is already low.
Thus, the CXB fluctuation is negligible.

The PIN signals, remaining after subtracting the NXB and CXB,
have statistical significance of 3 to 5 sigmas on individual observations,
and  8 sigmas when the three data sets are summed together.
Furthermore, the implied 12--20 keV signal rates, $\sim 0.02$ cts s$^{-1}$, 
amount to  6--10\% of the NXB count rate in the same band. 
This exceeds the typical NXB reproducibility during the present observations, 4\%,
as estimated in section~\ref{subsubsec:PIN_background}.
Therefore, the positive signals are thought to be significant
even considering the systematic  NXB errors.


After we had completed our data analysis, 
the background model {\it bgd\_D} was updated,
by taking into account dead-time correction 
and improving the estimation method (Mizuno et al., suzakumemo-2008-03).
Accordingly, we re-generated the background files using this improved method,
to find that it gives essentially the same results as the previous {\it bgd\_D}.
Therefore, we retain our original choice of {\it bgd\_A},
which tends to overestimate the NXB and hence gives more conservative results.
We later examine the effects of these different background models
by  artificially changing  the NXB  by $\pm 4\%$.   

\subsubsection{HXD light curves}
\label{subsubsec:PIN_lightcurve}

As described in section~\ref{subsec:XIS},
the XIS signal from the central region of M82 exhibited 
rather small (if any) variations during the present observations.
Supposing that the positive HXD-PIN signals,  as revealed above,
are also coming from an approximately identical region in  M82,
we expect rather constant count rates as well.
If, on the other hand, the positive HXD-PIN counts 
were an artifact of wrong NXB subtraction,
they would  vary with time typically on time scales of minutes to hours,
just reflecting those in the NXB \citep{kokubun2007}.
For this purpose, 
the 12--20 keV PIN light curves of the three observation were derived,
and presented in figure \ref{fig:lc_pin}
together with the behavior of the modeled NXB.

On all occasions,
the background-subtracted counts in figure~\ref{fig:lc_pin}
thus show  on-average positive values,
which agree with those in table \ref{tbl:cr_pin}.
Furthermore, the net signal (green) appears relatively constant within statistical errors,
in contrast to significant raw-count variations  (black)
that reflect those in the NXB (red). 
When fitted with a constant, these background-subtracted light curves
yield $\chi^2/{\rm d.o.f}$ of 41.7/28, 42.8/39, and 31.2/30, 
in the first, second, and third observations, respectively.
These imply  probabilities of  $\sim$ 5\%, 35\%, and 40\%
for the light curves to be consistent  with being constant.
Thus, we consider that  the positive HXD-PIN counts listed 
in table~\ref{tbl:cr_pin} persisted throughout the observations,
while the variable NXB was removed successfully.

In this way, we have carefully examined the HXD-PIN signals
using the Earth-occultation data and the on-source light curves.
The results confirm
that the positive HXD-PIN signals above the CXB are real,
and are  coming from the M82 sky region.

\subsubsection{HXD spectra}
\label{subusec:PIN_spectra}

Figure \ref{fig:spec_pin} shows the full-range PIN spectra (black)
from the three observations,
compared with the corresponding NXB models ({\it bgd\_A}; red).
The figure also shows the NXB-subtracted data (green), 
and the predicted  CXB spectrum (blue)
based on the {\rm HEAO-1} results \citep{boldt1987,gruber1999}.
Thus, after the NXB subtraction, 
positive signals remain up to $\sim 30$ keV in all three observations.
Furthermore, the signal  exceeds the expected CXB at least up to $\sim 20$ keV.
In the 12--20 keV band, these spectra imply 
an average signal count rate of  $\sim 0.023$ cts s$^{-1}$ above the CXB.
This value agrees with that derived from the simple count-rate
measurements (section~\ref{subsubsec:PIN_counts}; table \ref{tbl:cr_pin}),
and the light curve analysis  (section~\ref{subsubsec:PIN_lightcurve}).

\subsection{Spectral Model Fitting Analyses}
\label{subsec:fits}

In this subsection,
we  perform model fitting first to the XIS spectra,
and then to those of HXD-PIN,
both averaged over the 3 days.
Afterwards, the XIS and HXD-PIN spectra are jointly analyzed.

\subsubsection{Model fits to the XIS spectra}
\label{subsubsec:XIS_fits}
Among the three  observations,
the 3.2--10 keV XIS spectra differ to some extent in their normalization;
as inferred from figure~\ref{fig:lc_xis},
those of the 2nd and 3rd observations
are higher by  2.1\% and lower by 10.6\%, respectively,
than  that of the first occasion. 
However, after scaling  by these factors,
the three spectra agree with one another within their errors,
which are typically 5\% per channel under the present binning.
Therefore, we hereafter analyze the XIS spectra
summed over the 3 observations.

We jointly fitted the 3.2--10 keV XIS FI and BI spectra,
thus summed over the three observations,
with a model consisting of a PL for the continuum 
and two Gaussians for the Fe-K$_{\alpha}$ lines.
The  XIS response file  and the associated ancillary file
was generated with ``xisrmfgen'' and ``xissimarfgen'' tool,
respectively \citep{ishisaki2007}. 
We constrained the overall model normalization to be the same
 between the FI and BI detectors.
The hydrogen column density describing photoelectric absorption 
is here and hereafter fixed at $1.1 \times 10^{22}$ cm$^{-2}$, 
as determined with {\rm Chandra} \citep{kaaret2006}
under the least contamination by the diffuse emission:
this amount of absorption attenuates the continuum at 3.2 keV only by 10\%,
and if treated  as a free parameter,
the absorption would couple too strongly with the continuum slope.
The model yielded a fit as shown in figure \ref{fig:spec_xis} (left),
with the best-fit  parameters  summarized in table~\ref{tbl:spec_para} 
under the column of ``XIS-only".
Thus, the PL  photon index became $\Gamma = 1.67$.
Although not given in the table,
one Gaussian is centered at an energy of $6.67 \pm 0.02$ keV, 
with an intrinsic width (in Gaussian sigma) of $< 0.11$ keV
and an equivalent width (EW) of $59 \pm 11$ eV.
Its inclusion improved the fit by 
$\Delta \chi^2 \sim 88$.
The other Gaussian, with its energy fixed at 6.4 keV and assumed to be narrow,
exhibited an  EW of $19 \pm 8$ eV,
and improved the fit by $\Delta \chi^2 \sim 15$.

The 2--10 keV flux detected with the XIS, 
averaged over the three data sets,
becomes $2.1 \times 10^{-11}$ erg s$^{-1}$ cm$^{-2}$,
which yields a 2--10 keV luminosity of $3.4 \times 10^{40}$ erg s$^{-1}$ at 3.6 Mpc.
This is typical of, or relatively lower than,
those  measured previously from narrow regions 
(depending on the angular resolution) around X-1,
which scatter over $(1-6) \times 10^{-11}$ erg s$^{-1}$ cm$^{-2}$
(e.g.  \cite{griffiths1979,watson1984,fabbiano1988b,schaaf1989,
tsuru1992,tsuru1997,matsumoto1999,ptak1999, cappi1999,matsumoto2001,
kaaret2001,rephaeli2002,strohmayer2003,kaaret2006,dewangan2006}).

Although the above model approximately reproduces the observed XIS spectra,
the fit is formally unacceptable with $\chi^2/\nu = 1.83$ (table~\ref{tbl:spec_para}).
As shown in figure \ref{fig:spec_xis}a,
the model overpredicts the data above $\sim 7$ keV,
suggesting the spectrum to have 
a more convex shape than a simple PL.
Actually, the best-fit photon index becomes discrepant as 
$1.56 \pm 0.03$  and  $2.0 \pm 0.2$, 
when the fit energy range is restricted to 3.2--6.0 keV and 7--10 keV, respectively.
The PL fit could be made acceptable by 
allowing the low-energy absorption to vary freely,
but then the column density must become $\gtrsim 3.7 \times 10^{22}$  cm$^{-2}$,
much exceeding the  Chandra value of $1.1\times 10^{22}$  cm$^{-2}$ \citep{kaaret2006}.
Therefore, the intrinsic continuum is suggested to be curved.
To express such a convex shape,
we fixed again the absorption to $1.1\times 10^{22}$ cm$^{-2}$,
and replaced the PL continuum by a cutoff PL  model,
namely a PL  multiplied by an exponential cutoff factor.
Then, as shown in figure \ref{fig:spec_xis}b 
and described in table~\ref{tbl:spec_para},
the  fit has  become acceptable with $\chi^2/\nu = 1.06$,
yielding $\Gamma \sim 0.76$ and a cutoff energy of $E_{\rm cut}=5.8$ keV.
Therefore, we conclude that the spectrum
has a more convex shape than a PL.

\subsubsection{Model fits to the HXD-PIN spectra}
\label{subsubsec:PIN_fits}

We conducted similar model-fitting analyses 
to the background (NXB+CXB) subtracted
12--20 keV PIN  spectra, 
again summed over the three observations.
The fit utilized a response matrix at  the HXD nominal position, 
{\rm ae\_hxd\_pinhxnom\_20060814.rsp},
and an ancillary response file generated by the {\rm hxdarfgen} tool.
The latter takes into account the fact that  the HXD FOV center
is $\sim 5'.7$ offset from the M82 nucleus.
Then, a PL model gave a successful fit with $\chi^2/\nu=9.2/6$,
yielding a photon index of $\Gamma = 3.4^{+1.0}_{-0.8}$
(statistical 90\% confidence errors),
and a 12--20 keV flux of $4.4 \times 10^{-12}$ erg s$^{-1}$ cm$^{-2}$.
These become $\Gamma = 3.7^{+3.1}_{-1.2}$ 
and $1.9 \times 10^{-12}$ erg s$^{-1}$ cm$^{-2}$
if the NXB is artificially increased (over-subtracted) by 4\%,
while $\Gamma = 3.1^{+0.6}_{-0.6}$ 
and $6.7 \times 10^{-12}$ erg s$^{-1}$ cm$^{-2}$
when it is reduced (under-subtracted) by 4\%.
In any case, the implied spectral slope ($\Gamma = 3-4$) is significantly 
steeper than that  found in the XIS range ($\Gamma =1.67$),
reinforcing the gradual spectral steepening
found  in the XIS band.

\subsubsection{Joint fits to the XIS and PIN spectra}
\label{subsubsec:joint_fits}

In order to best utilize the wide-band capability of Suzaku,
we fitted the XIS and HXD spectra simultaneously,
using respective averages over the three observations.
Since the energy range becomes much expanded,
we tried  a larger variety of continuum models than employed so far.
In all modelings, the two Gaussians were included 
in the same manner as in section~\ref{subsubsec:XIS_fits}.
The model normalization was constrained to be the same
between the XIS and HXD-PIN,
because the HXD-PIN  ancillary response file (section~\ref{subsubsec:PIN_fits})
has already taken into account the different  
FOVs between the two instruments
(assuming that the HXD-PIN signal is coming from the M82 central region).
The obtained fits are shown in figure~\ref{fig:spec_xis+pin},
while the fit parameters are summarized in table \ref{tbl:spec_para}
under the column of  ``XIS+PIN".
(The table also gives  the results obtained without using the HXD data.)
Furthermore,  like in section~\ref{subsubsec:PIN_fits},
we considered the case where the NXB of HXD-PIN is
systematically  changed by $\pm4\%$.

First, we tried a PL model, 
but the joint fit was not acceptable with $\chi^2/\nu \geq 2.6$
even allowing for the 4\% NXB uncertainty (table~\ref{tbl:spec_para}).
As shown in figure \ref{fig:spec_xis+pin}a,
the model over-predicts the XIS and HXD-PIN data in the energy ranges of $\geq 7$ keV.
Although the fit is improved to $\chi^2/\nu \sim 1.9$ 
by tentatively allowing the model normalization 
to take separate values between the two instruments,
the HXD vs. XIS normalization ratio then became $\leq 0.7$.
This is too small,
considering the XIS vs. HXD calibration uncertainties 
which are at most $\sim 15\%$ \citep{kokubun2007}.
Thus, the joint fit  reinforces the inference made so far,
that the X-1 spectrum is more convex than a PL.
Below, we return to our original procedure of equalizing
 the model normalization between the two instruments.

Next, we tried an empirical thermal bremsstrahlung (bremss) model.
Although the fit is better  
($\chi^2/\nu = 1.3-2.2$ depending on the NXB systematic errors)  
than the PL fit, 
the model is still over-predicting in the $\geq 7$ keV energy range
as shown in figure \ref{fig:spec_xis+pin}b.
The obtained temperature, $T \sim 14$ keV, is higher 
than the values of $5-8$ keV measured with {\rm Ginga} \citep{tsuru1992}, 
{\rm BeppoSAX} \citep{cappi1999}, 
and {\rm RXTE} \citep{rephaeli2002}.
These results remain essentially unchanged even 
when changing the PIN NXB by $\pm 4\%$.

Third, we employed a cutoff PL model.
As shown in figure \ref{fig:spec_xis+pin}c and table \ref{tbl:spec_para},
it  has given $\chi^2/\nu = 1.07-1.26$,
together with  $\Gamma = 0.5-0.9$ 
and $E_{\rm cut} = 4-8$ keV.
Therefore, this model remains successful 
even including the HXD-PIN data.
Because a cutoff PL gives an approximation to an unsaturated thermal
Comptonization process of some soft seed photons,
we replaced it with a thermal Comptonization model,
called ``CompTT'' \citep{titarchuk1994} in \texttt{xspec}.
Since the input soft-photon energy is not well constrained by the present data,
we fixed it to several values between 0.1 and 0.3 keV.
Then, as presented in figure \ref{fig:spec_xis+pin}d,
the  fit was similarly acceptable with $\chi^2/\nu = 1.09-1.32$,
and gave an electron temperature  of $T_{\rm e} = 2.3-2.9$ keV
and an optical depth of $\tau = 7-9$;
the results were essentially independent of the soft-photon
temperature in the above-mentioned range.
The obtained values of   $T_{\rm e}$ and $\tau$ are both consistent
 with the {\rm XMM-Newton} results 
on M82 X-1 by \citet{agrawal2006}. 

Finally, to evaluate a scenario of  emission  from an optically-thick accretion disk,
we tested a ``disk blackbody" (\texttt{diskbb}) model 
which represents multi-color blackbody emission 
from a standard accretion disk \citep{mitsuda1984, makishima1986}.
As shown in figure \ref{fig:spec_xis+pin}e,
the fit was rather poor with $\chi^2/\nu =1.32-1.72$,
because the model has too convex a shape 
and hence under-predicts the data in the higher energy range.
To improve the fit, we then generalized the \texttt{diskbb} model 
 to ``variable-$p$" model 
\citep{mineshige1994,hirano1995,kubota+max2004,kubota2005,okajima2006,mizuno2007}.
It assumes that the disk temperature scales as $r^{-p}$, 
where $r$ is the radius and $p$ is a free parameter,
with  $p = 0.75$ corresponding to  \texttt{diskbb}. 
It gives a good approximation to emission from
a slim disk \citep{watarai2000,ebisawa2003}, 
which is expected to form under very high accretion rates.
The fit has then become acceptable 
with $\chi^2/\nu = 1.09-1.24$ (figure \ref{fig:spec_xis+pin}f),
and the obtained parameters,
namely the  inner-disk temperature $T_{\rm in} = 3.4-3.6$ keV 
and $p = 0.61-0.65$, are consistent with those of \citet{okajima2006}
who applied the the same model to the {\rm XMM-Newton} data of X-1
acquired at a similar luminosity to the present case.

From these studies, 
we have found three models that successfully and simultaneously 
reproduce the XIS and HXD-PIN data.
A cutoff PL model with $E_{\rm cut} \sim 5$ keV,
a CompTT model invoking a relatively cool and thick electron cloud,
and a variable-$p$ model with a very high disk temperature.
In all these cases, the two Gaussians were found to take
nearly the same parameters as found with the XIS 
 (section~\ref{subsubsec:XIS_fits}).
 
As  seen from table~\ref{tbl:spec_para},
the addition of the HXD-PIN data have  little affected
the best-fit parameters of these successful models,
with similar or somewhat reduced errors.
In contrast, their inclusion has allowed us to much 
more securely exclude the other unsuccessful models.
In this sense, the XIS and HXD-PIN data are 
fully consistent with each other.
When the NXB level is artificially changed by $\pm 4\%$,
the fit goodness of all the three successful models degrades  considerably.
This reinforces the correctness of the  NXB subtraction,
unless the detected signal has a very 
complex broad-band continuum shape
which requires a composite modeling.

\subsection{Estimation of Contamination from Other Sources}
\label{subsec:contami}

Although we have so far quantified the broad-band Suzaku spectra,
the results may not be totally ascribed to X-1,
because the HXD-PIN data are likely to be  contaminated  by M81,
as well as by other discrete sources in M82.
The latter must contaminate the XIS data as well,
because many of them are clustered in too narrow a region,
$\sim 30''$ in radius around the M82 center,
to be resolved with the XIS.

\subsubsection{M81}

As described in section~\ref{sec:observations} 
and illustrated in firgure~\ref{fig:image},
the  M81 nucleus  was  on the edge of the HXD-PIN FOV,
with  transmission of at most $1- 3\%$.
Therefore, the contamination is expected to be small,
but may not be totally neglected.

Using  {\rm BeppoSAX}, \citet{pellegrini2000} measured the 10--100 keV flux 
of  the M81 nucleus as $7.4 \times 10^{-11}$ erg s$^{-1}$ cm$^{-2}$,
and described its spectrum by a PL with $\Gamma \sim 1.84$.
The  12--20 keV flux is implied to be 
$1.5 \times 10^{-11}$ erg s$^{-1}$ cm$^{-2}$.
According to \citet{laparola2004}, furthermore, 
the M81 nucleus varied in the 0.5--2.4 keV band  
by a factor of 4 through past observations,
and became up to $\sim 1.4$ times 
as bright as observed by \citet{pellegrini2000}.

Adopting an extreme assumption
that the M81 nucleus was at its brightest level during the present observations,
and that the spectrum retained  $\Gamma \sim 1.8$,
the angular transmission described in section~\ref{sec:observations}
predicts this low-luminosity AGN  to contribute
up to  $\leq (2- 6) \times 10^{-13}$ erg s$^{-1}$ cm$^{-2}$
to the 12--20 keV HXD-PIN data.
These correspond to $5-15\%$ of the background-subtracted PIN flux
(section~\ref{subsubsec:PIN_fits}).
Taking an exposure-weighted mean over the three occasions,
the M81 contamination to the HXD-PIN signal 
is estimated  to be at most 10\%.
Since the PIN spectra in figure~\ref{fig:spec_xis+pin}
have statistical errors of  several tens percent per bin,
and since M81 may not have been in the brightest phase, 
we conclude that the effect of the M81 nucleus is negligible.
This is an advantage of the HXD with its tightly collimated FOV.

\subsubsection{Other discrete X-ray sources in M82}

Besides X-1,
the central region within $\sim 30''$ of the M82 nucleus 
harbors about a dozen X-ray point sources 
more luminous than $\sim 10^{38}$ erg s$^{-1}$
\citep{griffiths2000,matsumoto2001,kaaret2006,strickland2007}.
To study their  fluxes and spectra, 
and estimate their contamination to the present Suzaku results,
we re-visited the {\rm Chandra} archival data.
By 2007 August,
the central regions of M82 were observed six times with the Chandra ACIS
for exposures longer than 10 ks;
these are ObsID 361, 1302, 2933, 6097, 5644, and 6361.
Using the CIAO 3.4 software package and standard screening criteria,
we extracted point sources 
which are located within  $2'.9$ of X-1;
these can contaminate both  the present XIS and HXD spectra.
Referring to the ROSAT PSPC image,
any X-ray sources that are outside the XIS integration radius
but within the HXD-PIN FOV are considered negligible.

Through this analysis, 
we confirmed the report by \citet{kaaret2006}
that a second brightest source,
which they call X42.3+59,
is located at  $\sim 5''$ off X-1;
we hereafter call it X-2.
Across the six {\rm Chandra} observations,
the 2--10 keV flux of X-2 varied largely,
from less than $\sim 10^{-13}$ erg s$^{-1}$ cm$^{-2}$
to $6.9 \times 10^{-12}$ erg s$^{-1}$ cm$^{-2}$,
with the latter amounting to 1/3 of the 2--10 keV XIS signal flux.
As already reported by \citet{kaaret2006},
the 3--8 keV spectrum of X-2 at its brightest phase 
was described by a PL with  $\Gamma \sim 1.3$,
absorbed by a column of $N_{\rm H} = (2-3) \times 10^{22}$ cm$^{-2}$,
and a soft-excess component was required if lower energies are included.

Fainter X-ray sources, other than X-1 and X-2, were found to have
a 2--10 keV luminositiy of $ \lesssim 3 \times 10^{39}$ erg s$^{-1}$ each.
Among the six {\rm Chandra} observations,
their summed 2--10 keV  flux remained relatively  constant
at  $ (7-9) \times 10^{-12}$ erg s$^{-1}$ cm$^{-2}$,
which accounts for about 1/3 of the XIS signal.
Their summed 3--8 keV spectrum  was reproduced by a PL with  $\Gamma = 1.5-1.7$.

As shown so far,
we expect the fainter discrete sources to account
for about 1/3 of the detected XIS signal.
Furthermore, another 1/3 could be ascribed to X-2
if it is in the brightest phase.
Therefore, the soft X-ray flux actually attributable to X-1
is estimated to be 2/3 (if X-2 is very faint)
to 1/3 (if X-2 is brightest) of the detected 2--10 keV XIS signal
of $2.1 \times 10^{-11}$ erg s$^{-1}$ cm$^{-2}$
(section~\ref{subsubsec:XIS_fits}).
These contaminants are expected to
introduce a larger uncertainty to the HXD-PIN results,
because their hard X-ray spectra cannot be measured separately.
Below, we consider this issue assuming two extreme cases.

\subsubsection{Re-evaluation of the spectra considering contamination}
\label{subsubsec:reeval}

We first consider an extreme case 
where X-2 is faint  enough to be neglected,
and hence we need to consider only the  fainter sources.
If  their summed  spectrum determined with Chandra
(a PL with $\Gamma = 1.5-1.7$)
were extending into the hard X-ray range,
the implied counts at 30 keV would reach $\sim 10\%$  of the  HXD-PIN background,
which is well above the systematic background uncertainty.
Furthermore,  the 30-- 40 keV signals,
attainable with HXD-PIN in the 77 ks exposure (the three observations summed),
would have a statistical significance exceeding 6 sigmas.
Thus, we must have detected their signals at least up to $\sim 40$ keV,
as illustrated in figure~\ref{fig:spec_contami}a.
However, the actual 12--20 keV PIN spectrum was rather soft with $\Gamma  \sim 2.5$, 
and was undetectable above $\sim 20$ keV.
Therefore, the summed spectrum from the fainter sources 
should bend in energies above $\sim 7$ keV.
This is reasonable,
because these sources are considered to be mainly luminous low-mass X-ray binaries, 
of which the 2--20 keV spectrum is empirically approximated
by a thermal bremsstrahlung model with a  temperature of
$T = 7-13$ keV \citep{makishima1989}.
Such a bremsstrahlung model has a mildly curved shape,
and its  3--8 keV portion can be approximated 
by a PL with $\Gamma \sim 1.5$ \citep{makishima1989},
in agreement with the Chandra measurement.

With the above consideration,
we modeled the summed contribution of the faint sources
by a  $T = 10$ keV thermal bremsstrahlung model,
of which the 2--10 keV flux was fixed at $8 \times 10^{-12}$ erg s$^{-1}$ cm$^{-2}$
as specified by the  {\rm Chandra} data.
Including this  model as  a fixed component,
we jointly re-fitted the same {\rm Suzaku} XIS and PIN spectra
using the same family of models  as in section~\ref{subsubsec:joint_fits}.
As shown in table \ref{tbl:spec_contami_para} under the column  ``Case 1",
and presented in figure~\ref{fig:spec_contami}b,
the obtained results are close to those obtained before.
However, the inferred X-1 spectrum became  softer,
because the $T = 10$ keV thermal bremsstrahlung model,
which is now  subtracted away,
is somewhat harder than the original wide-band Suzaku spectrum.
As a result,  not only the successful three models,
but also the simple multi-color disk model with $T_{\rm in} \sim 3.1$ keV,
have become acceptable.
Equivalently, the variable-$p$ model gave $p \sim 0.75$,
which just implies a standard accretion disk.
After subtracting the contamination,
the 2--10 keV flux attributable to X-1 becomes $1.2 \times 10^{-11}$ erg s$^{-1}$ cm$^{-2}$,
and the luminosity therein is $1.9 \times 10^{40}$ erg s$^{-1}$.

As the other extreme, we may consider the case
where the present observations happened to catch X-2
at its brightest level ever recorded by {\rm Chandra},
and hence its contribution to the Suzaku spectra must be considered as well.
Although the X-2 spectrum obtained by {\rm Chandra} 
at the brightest epoch was rather hard ($\Gamma  \sim 1.3$),
it cannot keep extending into the $>10$ keV energy range,
for the same reason as discussed just above for the fainter sources.
Thus, the X-2 spectrum is inferred to have 
a power-law  shape in the 3--8 keV range,
together with a high-energy turn over, 
and a soft excess \citep{kaaret2006}.
Such a spectral composition is typical of  ``PL state" ULXs
(e.g., \cite{feng+kaaret2005,stobbart2006}).
Although these objects generally have a steeper range of spectra with $\Gamma= 1.6-2.5$, 
a case with $\Gamma \sim 1.4$ is known \citep{kubota2001b}.
Regarding X-2 as such an object,
we decided to express its spectrum by a cutoff PL model 
with $\Gamma=0.5$ and a cutoff energy of  $E_{\rm cut} = 5.0$ keV,
absorbed with $N_{\rm H} = 2.5 \times 10^{22}$ cm$^{-2}$
\citep{kaaret2006}.
The adopted cutoff energy  is a typical value found among 
power-law type ULXs \citep{miyawaki2008,mizuno2007},
and the photon index of 0.5  was chosen so that the model,
of the form $\propto \exp(-E/E_{\rm cut}) E^{-\Gamma}$, 
mimics the $\Gamma=1.3$ PL in the 3--8 keV band.
Its  normalization was  adjusted to reproduce  the {\rm Chandra} flux.

Including these two fixed models
(one  for X-2 and the other for the fainter sources),
we repeated the fitting procedure;
the obtained  results are presented in figure~\ref{fig:spec_contami}c,
and the parameters are given in table \ref{tbl:spec_contami_para}
under the column  denoted  ``Case 2".
Although the 2--10 keV flux attributable to X-1 further 
decreased to $7.0 \times 10^{-12}$ erg s$^{-1}$ cm$^{-2}$,
and its 2--10 keV  luminosity to $1.1 \times 10^{40}$ erg s$^{-1}$,
the spectral results have remained mostly the same 
as the case  of X-2 being negligible.
This is because the assumed X-2 spectrum  is
relatively similar in shape to the original Suzaku XIS plus HXD-PIN spectra,
as is clear from the cutoff PL fit parameters in table~\ref{tbl:spec_para}.

These studies confirm 
that the removal of the estimated contributions from X-2
(under the most extreme assumption)  and the other fainter sources
leaves significant 3--20 keV counts that are attributable to X-1.
These remaining hard X-ray photons still  form a clearly curved spectrum,
which excludes a PL and  a thermal bremsstrahlung modeling,
similar to the case of ignoring the contamination.
Thus, except for the normalization,
our basic results on X-1 are relatively unaffected by the contaminating sources,
in spite of apparently large uncertainties that  they introduce.
This is because we would encounter an unphysical 
condition of negative X-1 flux at $\sim 20$ keV 
if the contaminants had significantly harder spectra,
while we can simply utilize the parameters in table~\ref{tbl:spec_para} 
if  the spectra of the contaminants were much softer 
and hence negligible in the HXD-PIN range.


\section{Discussion}

\subsection{Summary of the results}
\label{sebsec:summary}

With the Suzaku XIS, we detected a 2--10 keV flux of 
$2.1 \times 10^{-11}$ erg s$^{-1}$ cm$^{-2}$ 
that  is attributable to the assembly of luminous point sources
within $3'$  of the nucleus of M82.
Of this flux, about 2/3 can be accounted for by the sum 
of the two most luminous  sources, X-1 and X-2,
whereas the remaining 1/3 is ascribed to the other fainter ones.
The share of 2/3 is totally attributable to X-1 
if the highly variable X-2 was negligibly dim at that time,
whereas it is  further subdivided roughly equally between X-1 and X-2 
if  X-2 was at the brightest state ever recorded.
We hence estimate  the 2--10 keV flux of X-1 
as $(7-14) \times 10^{-12}$ erg s$^{-1}$ cm$^{-2}$,
and its 2--10 luminosity as $(1.1-2.1) \times 10^{40}$ erg s$^{-1}$ at 3.6 Mpc,
both with a typical systematic uncertainty by a factor of two.
The estimated luminosity of X-1 is similar to,
or somewhat lower than,
those so far measured from this source.

From a wider sky region including M82,
we also detected significant hard X-ray emission  with the Suzaku HXD-PIN,
at a 12--20 keV flux of $4.4 \times 10^{-12}$ erg s$^{-1}$ cm$^{-2}$.
When summed over the three observations,
the HXD-PIN signal (after subtracting the NXB and CXB) 
has a statistical significance of 8 sigmas,
and about twice as high as the systematic background uncertainty.
Therefore, the signal is considered real.
The HXD data define a rather steep spectrum with $\Gamma = 3-4$.

The acquired HXD-PIN data are fully consistent with them
coming from the same sources as detected with the XIS.
Regardless of the way of removing the estimated contamination
from X-2 and the fainter sources (section \ref{subsec:contami}),
the overall Suzaku spectrum, 
spanning 3--20 keV (with a narrow gap over 10--12 keV),
is clearly more convex than a single PL,
even though it can be approximated by a PL over a limited 3--8 keV range.
This reconfirms previous broad-band studies  with Ginga \citep{tsuru1992}, 
BeppoSAX \citep{cappi1999}, and RXTE \citep{rephaeli2002},
which all  reported ``thermal" type spectral shapes 
rather than straight PL like ones.
Clearly,  relativistic jet emission, 
invoked by some authors (e.g., \cite{krauss2005}) to explain the behavior of some ULXs,
is inappropriate at least in the case of M82 X-1,
since the observed spectral curvature ($|\Delta \Gamma| \gtrsim 1$ across $\sim 10$ keV) 
is too strong to be attributable to synchrotron cooling or burn-off effects.

The curved spectrum of M82 X-1 can be expressed by 
either of the three successful continuum models;
an empirical cutoff PL model,
a thermal Comptonization  model invoking  a cool and thick electron cloud,
and an optically-thick emission from a rather hot accretion disk
(section \ref{subsec:contami}).
Details of their model parameters depend to some extent 
on the assumed broad-band spectra of the contaminants.
Superposed on the continuum, 
we also detected two iron emission lines.

\subsection{An interpretation of the curved spectrum}
\label{subsec:interpretation}

Although ULXs show a variety of spectral shapes,
two prototypical spectral states have been recognized through past observations
(e.g., \cite{makishima2000,sugiho2003,feng+kaaret2005,dewangan2006,
stobbart2006,mizuno2007,makishima2007a,makishima2007b,miyawaki2008}).
One is ``PL state",
wherein the spectrum below $\sim 10$ keV is approximated by a single PL model,
but is often  accompanied by a soft excess 
(e.g., \cite{miller2003,soria2007})
as well as  a mild turn-over toward higher energies 
(e.g., $>6$ keV; \cite{mizuno2007}).
We already argued in section~\ref{subsubsec:reeval}
that the second brightest source, M82 X-2,
 is likely to be an object in the PL state.
The other is ``convex-spectrum (hereafter CS) state" \citep{makishima2007a},
in which the spectrum exhibits a significantly curved shape 
even in the limited energy range of $\sim 1$ to $\sim 10$ keV,
so that an MCD model gives a better fit than a PL.
The two ULX states are not considered to represent distinct classes of objects,
since an increasing number of sources are confirmed  
to make transitions between these two spectral states 
(e.g., \cite{makishima2000,kubota2001b,sugiho2003,laparola2004,miyawaki2008}).

\citet{mizuno2007} and \citet{makishima2007a} 
utilized Suzaku data on NGC~1313,
to study spectral properties of two ULXs therein, X-1 and X-2.
Then, X-2 was found in the  CS state,
and its 0.5--10 keV spectrum was represented 
successfully by a variable-$p$ model with 
$p=0.63 \pm 0.03$ and $T_{\rm in} = 1.86 \pm 0.15$ keV.
In contrast, the other source, X-1, was in  the PL state, 
because  its spectrum was much less convex
(approximated by a PL with $\Gamma\sim 2.1$),
with a hint of soft excess and a mild high-energy turn over.
Thus, the XIS spectrum of NGC~1313 X-1
was reproduced successfully by  a cutoff PL continuum 
with  $E_{\rm cut} = 3.41^{+0.57}_{-0.40}$ keV and $\Gamma=0.89 \pm 0.20$,
plus a cool ($T_{\rm in} \sim 0.2$ keV) \texttt{diskbb} for the soft excess.
The source exhibited a rather high 0.4--10 keV luminosity 
of  $3.3 \times 10^{40}$ erg s$^{-1}$.

As clarified in the present study,
the Suzaku spectrum of M82 X-1 is approximated 
by a PL model with $\Gamma\sim 1.7$  in the limited 3--8 keV range,
while it is reproduced by a cutoff PL model with  $E_{\rm cut} \sim 4$ keV 
and $\Gamma = 0.2-0.3$ (table~\ref{tbl:spec_contami_para})
when the broad  3--20 keV band is incorporated.
These spectral properties of M82 X-1, as well as its luminosity,
are very similar to those of NGC~1313 X-1 as recorded with Suzaku,
although the values of $\Gamma$ differ by $\sim 0.5$.
Based on these similarities, 
we conclude that M82 X-1 was in the PL state during the present observations.
This also implies that M82 X-1 can be regarded as an ``ordinary" ULX,
rather than a more enigmatic  class of accreting source,
even though it is rather extreme in terms of the luminosity.

While the cutoff PL is a purely empirical model without physical meaning,
it approximates the process of  unsaturated thermal Comptonization
(section \ref{subsubsec:joint_fits}).
In fact, the 3--20 keV spectrum of M82 X-1 has also been
reproduced reasonably well by the \texttt{compTT} model
(table~\ref{tbl:spec_contami_para}),
invoking a rather low electron temperature of $\sim 2.3$ keV
and a large optical depth of $\sim 10$;
the condition is close to a nearly saturated Comptonization.
As mentioned in  section \ref{subsubsec:joint_fits},
the parameters derived here agree very well with those of  the same object
obtained by \citet{agrawal2006} using the 3--10 keV  {\rm XMM-Newton} EPIC data. 
The Comptonization modeling has also provided reasonable account 
for the spectra, typically in the 1--10 keV bandbass,
of other PL-state ULXs  \citep{kubota2001a,fiorito2004,
makishima2007a, makishima2007b,soria2007,miyawaki2008}.

A valuable guideline to the understanding of ULXs may be 
provided by the study of  Galactic black-hole binaries (BHBs) \citep{makishima2007a}.
So far, a fair number of BHBs have been observed 
to leave the canonical standard-disk state,
and enter  ``Very High" state \citep{miyamoto1991,mcclintock2006},
when  their luminosities approach the corresponding Eddington limits.
Then, the spectrum becomes dominated by a  PL-like continuum 
with $\Gamma\ge 2.4$,
plus a soft excess attributable to the underlying disk emission.
Such examples include 
GX339$-$4 \citep{miyamoto1991},
GS~$1124-68$ \citep{miyamoto1993},
GRO~J1655$-$40 \citep{kubota2001a,kobayashi2003},
XTE~J1550$-$564 \citep{kubota+done2004}, 
GRS~$1915+105$ \citep{reig2003,done2004},
and 4U~1630$-$47 \citep{abe2005}.
These Very-High state spectra of BHBs have been interpreted successfully 
by assuming that a cool standard accretion disk becomes
surrounded by a cloud of hot electrons, or a ``corona",
so that a significant fraction of the disk photons get Comptonized by the corona
(e.g., \cite{done2007} and references therein).
According to \citet{kubota2001a},
the RXTE PCA spectrum of GRO~J1655$-$40 in the Very High state
can be understood by invoking a corona with
an electron temperature of $T_{\rm e} \sim 10$ keV
and a Compton optical depth $\tau \sim 2$,
located above a cool disk which supplies the seed photons.
\citet{kobayashi2003} incorporated}
the RXTE  HEXTE data of the same object
and derived $T_{\rm e}=75-85$ keV and $\tau=0.45-0.65$.
Although the solutions thus degenerate,
the Compton $y$-parameter becomes in either case $y \sim 0.3$.

After the initial suggestion by \citet{kubota2002}
and subsequent discussion by several authors
\citep{tsunoda2006,mizuno2007,makishima2007a,miyawaki2008},
we presume that the PL state of ULXs is
analogous to the Very High state of BHBs,
and that both appear at near the Eddington luminosity.
The bolometric  luminosity of M82 X-1 during the present
observation is estimated to be $L_{\rm bol} = (1.5-3) \times 10^{40}$ erg s$^{-1}$,
by integrating the cutoffPL model.
The result does not differ significantly if we instead use  the MCD fit.
Then, its identification with the Eddington limit 
leads to a mass estimate as $100-200~M_\odot$.
The  mass could be even  higher by a few times,
if considering that the highest recorded luminosity 
reaches  $\sim 1 \times 10^{41}$ erg s$^{-1}$,
and that the Very High state may appear at a
luminosity somewhat below the Eddington value
\citep{kubota+max2004}.
A possible formation scenario of such an intermediate-mass BH,
particularly in close vicinities of galaxy nuclei,
has been proposed by \citet{ebisu2001} and \citet{ebisu2006}.
We admit, on the other hand,
that the reality of such a cool and thick Compton cloud 
needs a careful consideration,
from the viewpoints of its stability, energetics, 
and geometry with respect to the seed photon source.
These issues are discussed by  \citet{agrawal2006}.

\subsection{Another interpretation of the spectrum}
\label{subsec:interpretation2}

Assuming that the PL state of ULXs 
is analogous to the Very High state of BHBs,
we may identify, after some previous works
\citep{mizuno2001,watarai2001,tsunoda2006,makishima2007a},
the CS state of ULXs to the slim-disk state of BHBs
that is observed at their highest luminosities 
\citep{kubota+max2004,abe2005,makishima2007b}.
This analogy has the following three  grounds. 
First, a ULX generally becomes more luminous, 
by a factor of 1.5--3 (in the 1--10 keV band),
when it makes a transition from the PL state to the CS state
\citep{kubota2001b,laparola2004,sugiho2003,miyawaki2008}.
This luminosity relation, including its sense and magnitude,
is similar to that observed between the Very High state and the slim-disk state of BHBs.
Second, the CS-state spectra of ULXs can be generally reproduced successfully by 
the variable-$p$ model with $p\sim 0.6$ (e.g., \cite{tsunoda2006,mizuno2007}).
Finally,  when multiple observations of a single CS-state ULX are 
analyzed by a simple MCD model,
we observe the luminosity to scale as  $\propto T_{\rm in}^2$ \citep{mizuno2001}.
The second and third properties agree with
the observed behavior of BHBs in their slim-disk state \citep{kubota+max2004},
as well as predictions by the slim-disk theory (e.g., \cite{watarai2001,kawaguchi2003}).

In practice, the above state classification scheme 
of ULXs can be sometimes rather ambiguous,
because  a  variable-$p$ spectrum with a very high disk temperature
takes a rather similar shape to a cutoff PL model.
For example, \citet{kiki2006}  applied the variable-$p$ modeling
to several PL-state ULXs observed with XMM-Newton.
The present M82 X-1 data are subject to the same  ambiguity,
because the cutoff PL model and the variable-$p$ 
model have been similarly successful.
In fact, \citet{okajima2006} interpreted an XMM-Newton spectrum of this object
(plus some other point sources inclusive)
as arising from an extreme slim disk with a highly super-Eddington luminosity,
formed around a BH of  $20-30~M_\odot$.
Even utilizing the broader energy band realized by Suzaku,
the degeneracy was not solved,
and it remained  after removing the source contamination.

Let us briefly examine the slim-disk interpretation  
of the present Suzaku data for its consistency.
The variable-$p$ fit to the XIS plus HXD data gave extremely high disk
temperatures  as $T_{\rm in} =3.0-3.5$ keV 
(table~\ref{tbl:spec_para} and table~\ref{tbl:spec_contami_para}), 
regardless of the treatment of the source contamination.
It significantly exceeds those observed from other PL-state ULXs, 
including  $1.3-1.9$ keV  from NGC 1313 X-2 at a 0.4--10 keV luminosity of
 $0.6 \times 10^{40}$ erg s$^{-1}$ \citep{mizuno2007},
and $1.5-1.7$ keV  from M81 X-9 
at a 0.5--10 keV luminosity of $2.4 \times 10^{40}$ erg s$^{-1}$ \citep{tsunoda2006}.
On the other hand, the  temperature index of M82 X-1
 was obtained as $p=0.64 \pm 0.02$ in table~\ref{tbl:spec_para},
or  $p=0.75$ in  table~\ref{tbl:spec_contami_para}.
Since the latter is consistent with the condition of a standard disk,
we may calculate its  innermost disk radius $R_{\rm in}$
via $L_{\rm bol} = 4 \pi (R_{\rm in}/\xi)^2 \sigma T_{\rm in}^4$;
here $\sigma$ is the Stefan-Boltzmann constant,
$\xi \sim 1.2$ is a correction factor in \citet{makishima2000},
and the disk inclination is assumed for simplicity as $\sim 60^\circ$.

Taking $T_{\rm in} = 3.0$ keV (table-\ref{tbl:spec_contami_para})
and $L_{\rm bol} =(1.5-3) \times 10^{40}$ erg s$^{-1}$ (section~\ref{subsec:interpretation}),
the above procedure yields  $R_{\rm in} =37 -52$ km,
which corresponds to the last stable orbit of a 
non-rotating BH with a mass of only $4-6~M_\odot$.
Then, X-1 would be inferred to be shining
at  25-35 times the Eddington luminosity.
According to a recent numerical study by \citet{kiki2008},
this mass estimate (and hence that of the super-Eddington factor) 
would not differ by more than a factor of 2,
even if the system is in a slim-disk condition.
Therefore, we must conclude the source to be in a very deep slim-disk condition.
However,  this conclusion is not easily reconciled with the measured values of $p$,
which is closer to the standard-disk condition of 0.75
rather than to the limiting slim-disk condition of 0.5.
As far as non-rotating BHs are considered,
we hence regard the slim-disk interpretation as not being self-consistent;
the observed spectrum is simply  too hard to be regarded
as emission from an optically-thick accretion disk.

If the BH in M82 X-1 is spinning  rapidly,
its observed properties  could be understood 
more easily in terms of standard accretion disks.
This is because the last stable orbit would  be then  reduced down to 
$\sim 1/6$ times that of a non-spinning BH of the same mass.
This idea was first proposed by \citet{makishima2000}  
to explain CS-state spectra of ULXs,
and recently applied successfully by \citet{isobe2008} to a transient ULX.
If M82 indeed hosts a maximally rotating BH,
the BH mass estimated from the observed $R_{\rm in}$
would increase to $24-36~M_\odot$.
However, we are still left with a super-Eddington factor of 4--9,
and have to take into account various relativistic effects \citep{ebisawa2003}.
Therefore, the physical reality of this interpretation is open.
In addition, the reported spectral hardening toward lower fluxes (of presumably X-1 itself), 
based on the RXTE monitoring \citep{rephaeli2002},
is opposite to the behavior of CS-state ULXs \citep{mizuno2007}.
From these considerations,
we regard our baseline interpretation presented
in section \ref{subsec:interpretation} as more appropriate.

Incidentally, the state classification scheme of BHBs employed above 
is based empirically on the observed {\it luminosity},
but a more physically reasonable scheme should be based
on the {\it accretion rate normalized to the Eddington limit.}
Considering that the radiative efficiency itself varies,
it is not obvious which of the two states,
the Very High state and the slim-disk state,
correspond to higher normalized accretion rates.
They might even have nearly the same accretion rate,
and are controlled by some other parameters including hysteresis effects.
In GRO~J1655$-$40 \citep{kubota+max2004}, in fact,
the system was observed to evolve directly 
from the standard-disk  state into the slim-disk state
on one occasion when the luminosity was increasing,
while from the slim-disk state to the Very High state
and then back to the standard-disk state 
on two occasions during the luminosity descent.
A similar hysteresis effect,  though less clear,
is suggested  in the case of 4U~1630$-$47 \citep{abe2005}.
Clearly, we need deeper understanding of these two high-luminosity states
that have been recognized relatively recently.

\subsection{Subsidiary Spectral Features}
\label{subsec:iron_features}

\subsubsection{Possible iron edge}

Even with the successful three continuum models found in 
section~\ref{subsubsec:joint_fits},
we observe some data deficits in the $7-9$ keV energy range
 (figure~\ref{fig:spec_xis+pin}).
To account for this structure,
we  multiplied an absorption edge factor to the continuum models
fitted to the XIS and HXD spectra.
Regardless of the choice among the three successful continua,
the value of $\chi^2$ then decreases by $\sim 15$
for a decrease of degree of freedom by 2,
implying that the edge factor  improves the fit with  $>99\%$  significance.
The edge energy was  constrained as $7.1 \pm 0.1$ keV,
which agrees with the K-edge energy of nearly neutral iron.
Its optical depth was obtained as  $0.06-0.13$, 
depending on the  continuum models and the NXB systematic errors.

The suggested excess Fe-K edge could be reconciled
with the adopted low-energy absorption ($ 1.1 \times 10^{22}$ cm$^{-2}$)
if the  iron to alpha-element abundance ratio were
5 to 10 times higher than the solar ratio.
However, this is opposite to the measured abundance pattern
of X-ray emitting materials in M82 \citep{ranalli2008},
and hence unlikely.
An alternative possibility is
that X-1 is in fact rather heavily absorbed by a column 
up to $ (4.5-9) \times 10^{22}$ cm$^{-2}$ (assuming solar ratios),
but the associated strong  low-energy absorption is
masked by separate softer components.

Yet another possibility is
that the spectrum bears a ``reflection" component from cold materials,
and the Fe-K edge feature imprinted in that component is being observed.
To examine this scenario,
we removed the Fe-K edge factor from the cutoff PL continuum,
and added to it a reflection component, \texttt{pexrav} \citep{pexrav1995},
with the reflector iron abundance fixed at 0.6 solar,
those of lighter elements at 1.2 solar \citep{ranalli2008},
and the inclination at $60^\circ$.
We used the cutoff PL continuum itself as the input to the reflection.
Then, the fit chi-squared decreased (from the original cutoff PL modeling) by $\sim 14$,
for a decrease of degree of freedom by 1.
Therefore, the improvement is comparable to that obtained by applying the Fe-K edge.
The reflector solid angle $\Omega$, as seen from the nucleus,
is suggested to be relatively large as $\Omega/2\pi \sim 1$,
but with a considerable uncertainty,
as the associated  Compton hump is not well constrained by the HXD data.
We consider this scenario most likely to be the case.


\subsubsection{Iron emission lines}

Let us also discuss the two Fe-K lines detected with the XIS (figure~\ref{fig:spec_xis}).
Using  Chandra and XMM-Newton, 
\citet{strickland2007}  showed 
that both these lines are associated primarily to diffuse emission,
with insignificant contributions from bright point sources including X-1.
Over a circular region of radius $29''$ (0.53 kpc) around the nucleus, 
they measured the 6.7 keV and 6.4 keV components to have photon fluxes of 
 $(6.9 \pm 4.6) \times 10^{-6}$ ph cm$^{-2}$ s$^{-1}$
and $(4.8 \pm 4.0) \times 10^{-6}$ ph cm$^{-2}$ s$^{-1}$, respectively 
(from their table~3).

While our data integration radius ($3'$) is significantly larger 
than that of \citet{strickland2007},
diffuse emission in the 2--8 keV band is mostly contained
within the $30''$ radius (figure 4f of \cite{strickland2007}).
Therefore, the Chandra and Suzaku results may be directly compared
to the first approximation.
From our analysis conducted in section~\ref{subsubsec:XIS_fits},
the stronger 6.7 keV line has a flux of 
$10.2 \pm 2.1$ while the weaker 6.4 keV line $4.5 \pm 2.0$,
both in units of $10^{-6}$ ph cm$^{-2}$ s$^{-1}$.
Thus, the Suzaku and Chandra results agree with each other
within respective errors.
The observed Fe-K line photons  are hence
attributed primarily to the diffuse component,
but some fraction of the 6.4 keV photons may be
associated with the putative reflection component of X-1.

Intriguingly, the co-existing 6.4 keV and 6.7 keV Fe-K lines have also 
been detected in the diffuse X-ray emission around our Galactic center 
(\cite{koyamaGC2007}, and references therein)
and along the Galactic ridge  \citep{ebisawa2008}.
Therefore, we may be observing a similar phenomenon from the M82 central region.
However, the emission in our Galactic center is
confined within a considerably narrower region of $\lesssim 150$ pc,
and its Fe-K photon luminosity, as measured with Suzaku,
is 3 orders of magnitude lower than that  from the central region of M82
(Tsuru, Nobukawa, and Koyama, a private communication).
Presumably, the diffuse emission in M82 is significantly 
enhanced due to the extreme star-formation activity.
Further discussion on the iron emission lines is beyond the scope of the present paper.

\subsection{Interpretations Invoking Anisotropic Emission}

In addition to the interpretations of ULXs 
assuming either  intermeddate-mass BHs (section \ref{subsec:interpretation})
or highly super-Eddington luminosities (section \ref{subsec:interpretation2}),
a third explanation of these sources invokes 
strong emission anisotropy enhanced toward us (e.g., \cite{king2002}).
Among such ideas, 
we have already argued in section~\ref{sebsec:summary}
against the relativistic beaming scenario.
The remaining possibility is geometrical collimation 
of radiation by a funnel of thick materials \citep{king2002}.
In such cases, we should expect strong spectral signatures due to radiation reprocessing,
including Fe-K edge features and fluorescent Fe-K lines,
like in some extreme Seyfert 2 galaxies (e.g., \cite{ueda2007}).
Indeed, we found the evidence of reflection by 
nearly neutral materials (section \ref{subsec:iron_features}).
However,  the observed  reprocessor solid angle,  $\Omega/2\pi \sim 1$,
and the equivalent width of the neutral Fe-K emission line,
$\sim 20$ eV (section~\ref{subsubsec:XIS_fits}),
are both far smaller than would be expected 
($\Omega/2\pi >> 1$, and an equivalent width of $> 1$ keV)
when the observed radiation is dominated by the reflected component.
Furthermore, no adequate explanation based on the collimation scenario
has been presented, 
either for the two types of  ULX spectra,
or for the transitions between them.
Therefore, the anisotropic radiation interpretation is unlikely
to give a successful account of the ULX phenomenon in general.

\section{Conclusion}

Based on our discussion, we conclude that 
M82 X-1 was in the PL state during the present Suzaku observation,
and was radiating roughly close to, or slightly below, the Eddington luminosity.
This leads to a mass estimate of  at least $100-200~M_\odot$,
which is in a good agreement with an independent estimate
based on the QPO frequency (section~\ref{sec:intro}).
We therefore consider M82 X-1 as a genuine intermediate-mass BH,
in the sense that it is significantly more massive than those BHs 
that can be explained by the standard stellar evolutionary scenario.


\clearpage

\normalsize
\onecolumn
\begin{table}[htbp]
  \caption{Signal and background count rates from HXD-PIN.}
  \label{tbl:cr_pin}
  \begin{center}
    \begin{tabular}{lcccc}
      \hline
      \hline
      & October 4 & October 19 & October 27 & Average \\
      \hline
      On-source & $0.295 \pm 0.006$ & $0.325 \pm 0.005$ & $0.343 \pm 0.006$ & --- \\
      NXB\footnotemark[$*$] & 0.260 & 0.280 & 0.300 & --- \\
      CXB\footnotemark[$\dagger$] & 0.018 & 0.018 & 0.018 & --- \\
      Net signal\footnotemark[$\ddagger$] & $0.016 \pm 0.006$ & $0.027 \pm 0.005$ & $0.025 \pm 0.006$ & $0.023 \pm 0.003$ \\
      \hline
      \hline
     \end{tabular}
  \end{center}
            \footnotemark[$*$] Modeled with {\it bgd\_A}.\\
            \footnotemark[$\dagger$] Predicted from the HEAO-1 measurements \citep{boldt1987,gruber1999}.\\
             \footnotemark[$\ddagger$] Net count rate, 
            obtained by subtracting the NXB and CXB from the on-source rate.
\end{table}

\begin{table}[htbp]
  \caption{Best-fit model spectral parameters with  90\% confidence errors.\footnotemark[$*$]}
  \label{tbl:spec_para}
  \begin{center}
    \begin{tabular}{lccccc}
      \hline
      \hline
      Models & Parameters  & XIS-only\footnotemark[$\dagger$] 
                                                                    & XIS+PIN\footnotemark[$\ddagger$] 
                                                                                          & NXB+4\%\footnotemark[$\S$] 
                                                                                                              & NXB-4\%\footnotemark[$\S$] \\
      \hline
      Power-law   
             &  $\Gamma$     & $1.67\pm 0.02$ 
                                                                      & $1.75\pm 0.02$
                                                                                         & $1.77\pm 0.02$
                                                                                                              & $1.73\pm 0.02$ \\
              &  $\chi^2/{\rm d.o.f}$ & 318.6/174 & 606.8/182 & 809.1/182 & 464.9/182 \\
      \hline
     Bremss  
            & $T$ (keV)           & $14.8^{+0.8}_{-0.7}$ 
                                                                     & $13.6^{+0.6}_{-1.0}$ 
                                                                                         & $13.6^{+0.8}_{-0.4}$ 
                                                                                                             & $13.5^{+0.8}_{-0.5}$ \\
             &  $\chi^2/{\rm d.o.f}$ & 223.9/174 & 292.2/182 & 408.3/182 & 242.0/182 \\
      \hline
      Cutoffpl    
              &  $\Gamma$      & $0.76^{+0.13}_{-0.15}$ 
                                                                     & $0.77^{+0.08}_{-0.15}$ 
                                                                                        & $0.55^{+0.08}_{-0.17}$ 
                                                                                                           & $0.89^{+0.13}_{-0.10}$ \\
              & $E_{\rm cut}$ (keV)& $5.8^{+1.0}_{-0.8}$
                                                                    & $5.7^{+0.5}_{-0.7}$
                                                                                         & $4.7^{+0.4}_{-0.6}$ 
                                                                                                             & $6.8^{+1.5}_{-0.8}$ \\
              & $\chi^2/{\rm d.o.f}$ & 183.7/173 & 193.6/181 & 228.6/181 & 202.1/181 \\
      \hline
     CompTT  
               &$T_{\rm e}$ (keV) & $2.5 \pm 0.1$ 
                                                                   & $2.5 \pm 0.1$
                                                                                        & $2.4 \pm 0.1$ 
                                                                                                             & $2.8 \pm 0.1$ \\
               &$\tau$              & $8.3 \pm 0.4$ 
                                                                  & $8.0\pm 0.4$ 
                                                                                        & $8.6\pm 0.4$ 
                                                                                                             & $7.4\pm 0.4$ \\
                &$\chi^2/{\rm d.o.f}$ & 184.5/173 & 197.2/181 & 203.5/181 & 239.4/181 \\
      \hline
     Diskbb 
                 &    $T_{\rm in}$ (keV)   & $2.77 \pm 0.04$ 
                                                                & $2.79 \pm 0.04$ 
                                                                                       & $2.75 \pm 0.04$ 
                                                                                                           & $2.80 \pm 0.04$ \\
               & $\chi^2/{\rm d.o.f}$ & 224.8/174 & 243.4/182 & 239.8/182 & 312.8/182 \\
      \hline
     Variable-$p$      
              & $T_{\rm in}$ (keV)   & $3.38^{+0.32}_{-0.20}$ 
                                                               & $3.58^{+0.28}_{-0.18}$ 
                                                                                     & $3.47^{+0.22}_{-0.07}$ 
                                                                                                         & $3.53^{+0.50}_{-0.21}$ \\
               &  $p$       & $0.64 \pm 0.02$ 
                                                              & $0.63\pm 0.02$ 
                                                                                   & $0.63^\pm 0.02$
                                                                                                        & $0.63\pm 0.02$ \\
            & $\chi^2/{\rm d.o.f}$ & 183.4/173 & 191.9/181 & 225.1/181 & 220.0/181 \\
      \hline
     \end{tabular}
  \end{center}
          \footnotemark[$*$] The absorbing column density is fixed at $1.1 \times 10^{22}$ cm$^{-2}$.\\
         \footnotemark[$\dagger$] Using the XIS FI and BI spectra in the 3.2--10 keV band.\\
         \footnotemark[$\ddagger$] From the  XIS and HXD-PIN joint fit.\\
         \footnotemark[$\S$] When the modeled NXB spectra are systematically over- or under-estimated by 4\%.
\end{table}

\begin{table}[htbp]
  \caption{Same as table \ref{tbl:spec_para}, but considering contamination from surrounding sources}
  \label{tbl:spec_contami_para}
  \begin{center}
    \begin{tabular}{lccc}
      \hline
      \hline
      Models  &  Parameters & Case 1\footnotemark[$*$] &Case 2\footnotemark[$\dagger$] \\
      \hline
     Power-law  &    $\Gamma$          & $1.64\pm 0.03$  & $1.71 \pm 0.04$ \\
                       & $\chi^2/{\rm d.o.f}$ & 538.0/182                       & 303.9/182 \\
      \hline
      Bremss     &   $T$ (keV)             & $18.2 \pm 1.1$       & $13.5^{+2.4}_{-0.9}$ \\
                       &$\chi^2/{\rm d.o.f}$  & 384.4/182                        & 239.0/182 \\
      \hline
      Cutoffpl    &   $\Gamma$            & $0.27^{+0.02}_{-0.04}$   & $0.17^{+0.06}_{-0.04}$ \\
                      & $E_{\rm cut}$ (keV)  & $4.2 \pm 0.1$        & $3.7\pm 0.1$ \\
                      & $\chi^2/{\rm d.o.f}$   & 193.2/181                       & 193.1/181 \\
      \hline
      CompTT   &   $T_{\rm e}$ (keV)   & $2.30 \pm 0.03$ & $2.33^{+0.04}_{-0.08}$ \\
                      & $\tau$                       & $9.7\pm 0.2$        & $9.2^{+0.2}_{-0.3}$ \\
                      & $\chi^2/{\rm d.o.f}$    & 196.8/181                      & 199.9/181 \\
      \hline
      Diskbb      &    $T_{\rm in}$ (keV)  & $3.09 \pm 0.07$ & $2.95^{+0.07}_{-0.10}$ \\
                      & $\chi^2/{\rm d.o.f}$    & 192.7/182                      & 195.0/182 \\
      \hline
      Variable-$p$  &  $T_{\rm in}$ (keV) & $3.09 \pm 0.07$ & $2.95^{+0.07}_{-0.11}$ \\
                          &$p$                           & $0.75^{+0.01}_{-0.02}$ & $0.75^{+0.02}_{-0.03}$ \\
                         &  $\chi^2/{\rm d.o.f}$  & 192.7/181                       & 194.9/181 \\
      \hline
      Flux          \footnotemark[$\ddagger$]&      (2--10 keV)           &        1.2               & 0.7            \\
       \hline
     \end{tabular}
  \end{center}
      \footnotemark[$*$] The case when spectra of the fainter contaminant sources (except  X-2) were removed.\\
     \footnotemark[$\dagger$] The case when spectra of both the  fainter sources and X-2 were removed.\\
     \footnotemark[$\ddagger$] In  units of $10^{-11}$ erg s$^{-1}$ cm$^{-2}$.
        Depends no more than 10\% on the spectral models.\\
  \end{table}

\clearpage

\bigskip

\begin{figure}[htbp]
  \begin{center}
    \FigureFile(160mm,){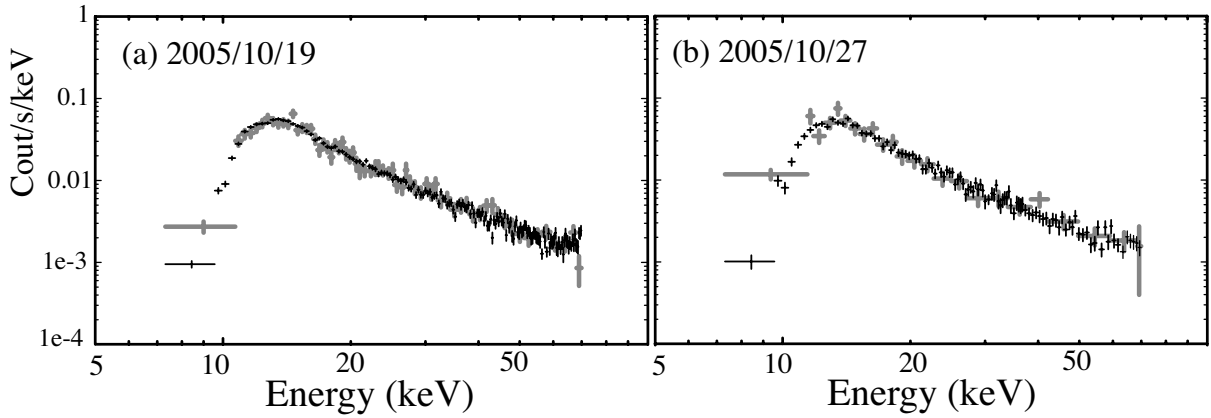}
  \end{center}
  \caption{
    Earth-occultation spectra of HXD-PIN (black) 
    obtained on (a) October 19 and (b) October 27,
    compared with the modeled non X-ray background (gray) to be observed 
    during the same time periods.
  }
  \label{fig:spec_pin_earth}
\end{figure}

\vspace*{-5mm}
\begin{figure}[htbp]
  \begin{center}
    \FigureFile(16cmm,){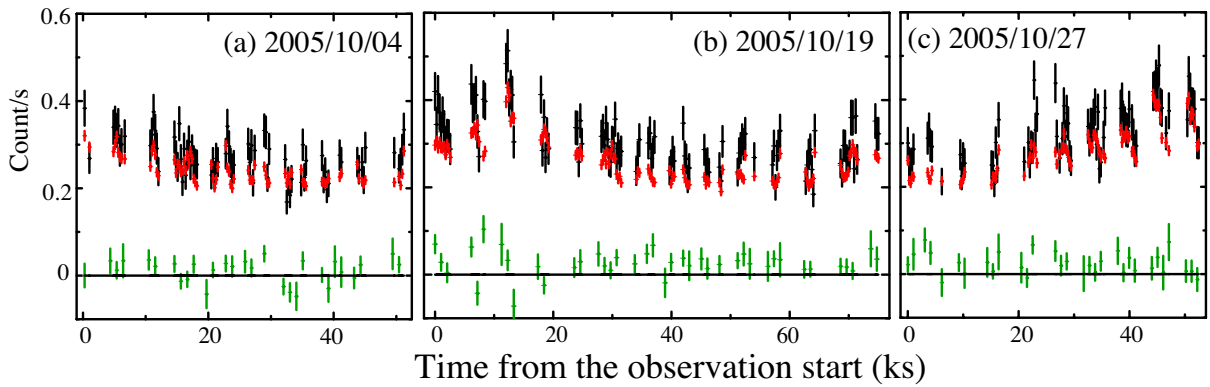}
  \end{center}
  \caption{
    The 12--20 keV light curves of the M82 region acquired with HXD-PIN
    on (a) October 4, (b) October 19, and (c) October 27.
    The observed raw data are shown in black, 
     while the modeled NXB data in red,  both using a binning of 256 s.
     Green data points show the signals remaining after subtracting 
     the NXB and the expected CXB, binned into 1024 s.
  }
  \label{fig:lc_pin}
\end{figure}

\begin{figure}[htbp]
  \begin{center}
    \FigureFile(16cmm,){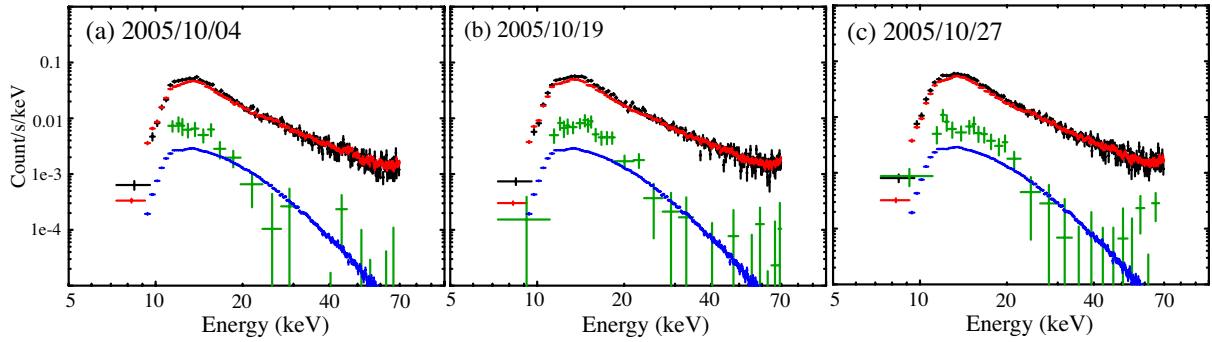}
  \end{center}
  \caption{
    Time-averaged HXD-PIN spectra obtained on (a) October 4, (b) October 19,
     and (c) October 27.
    Each panel shows the observed spectrum (black), the modeled NXB spectrum (red), 
    the signals remaining after subtracting the NXB (green), and the expected CXB spectrum (blue).
  }
  \label{fig:spec_pin}
\end{figure}

\begin{figure}[htbp]
  \begin{center}
    \FigureFile(16cmm,){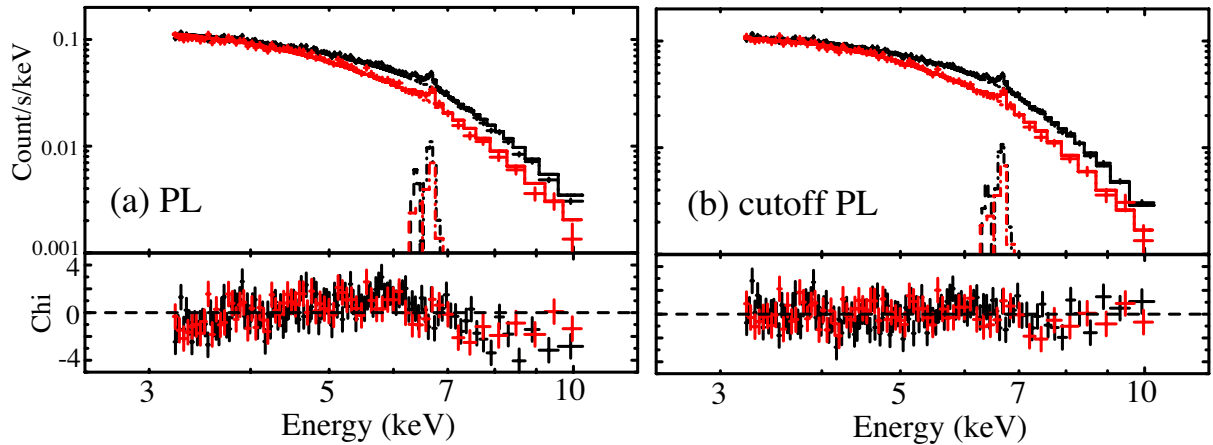}
  \end{center}
  \caption{(a) A 3.2--10 keV portion of the XIS FI (black) and BI (red)  
    spectra in figure~\ref{fig:spec_xis_wide}, presented together with the best-fit 
    absorbed power-law plus gaussian model (upper panel) and the fit residuals (lower panel).
    (b) The same as panel a, but the power-law is replaced by a cutoff power-law.
    }
  \label{fig:spec_xis}
\end{figure}

\begin{figure}[htbp]
  \begin{center}
    \FigureFile(145mm,){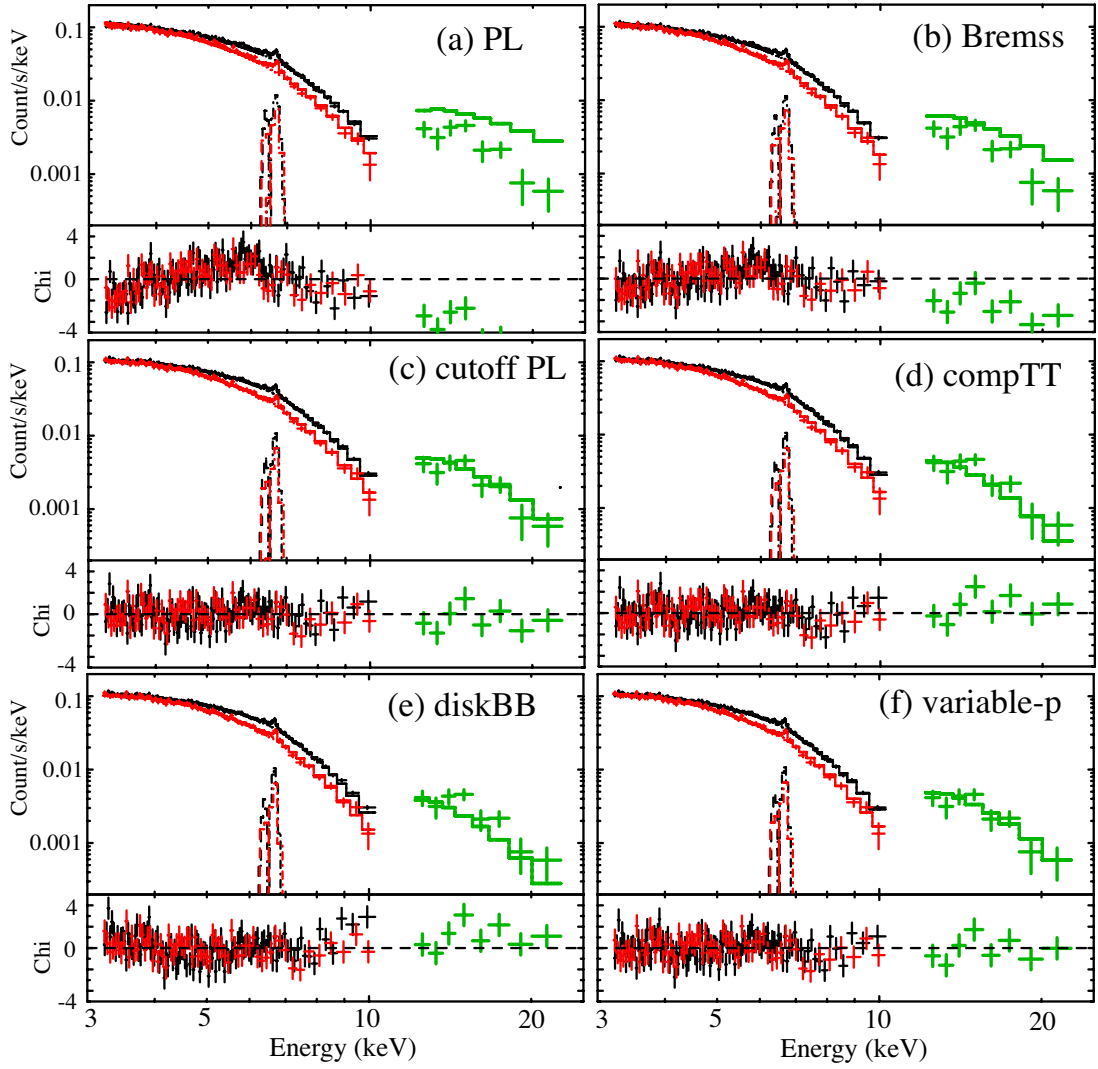}
  \end{center}
  \caption{
    The same as figure \ref{fig:spec_xis},
     but the fit jointly incorporates  the HXD-PIN spectrum (green)
    from which  the NXB (nominal) and CXB are both subtracted.
    The employed fitting models are; (a) power-law,  (b) bremsstrahlung, (c) cutoff power-law,
    (d) \texttt{compTT}, (e) \texttt{diskbb}, and (f) variable-$p$ disk model.
  }
  \label{fig:spec_xis+pin}
\end{figure}

\begin{figure}[htbp]
  \begin{center}
    \FigureFile(80mm,){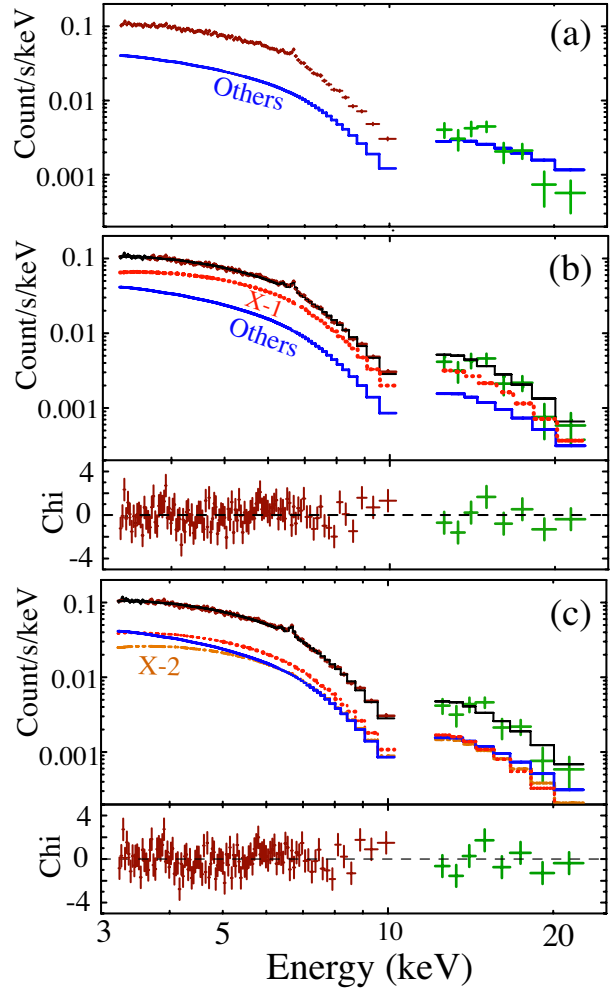}
  \end{center}
  \caption{ 
    The same joint XIS plus HXD fits as figure \ref{fig:spec_xis+pin},
     but incorporating estimated contributions from the contaminating sources
     (table~\ref{tbl:spec_contami_para}).
     The data points from XIS-FI and HXD-PIN are shown in brown and green,
     respectively, while the XIS-BI data are omitted for clarity.
     The two Gaussians, though included, are omitted for the same reason.
     (a) The case when the summed spectrum of ``the other fainter sources", 
     estimated in the XIS band (blue), extends into the HXD-PIN range with a PL shape.
     (b) The case (Case 1) when the spectrum of the fainter sources (blue) is approximated by a
      10 keV bremss model, with its normalization fixed in the XIS range. 
      The X-1 spectrum (red) is modeled by a cutoff PL with a free shape and free normalization.
      The total model prediction is indicated in black,
       and the fit residuals are shown in the lower panel.
      (c) The same as panel b, but when the X-2 contribution is included 
      as a fixed model (orange) represented by another cutoff PL model (Case 2; see text).
      In the HXD-PIN range, the three constituent components nearly overlap.}
  \label{fig:spec_contami}
\end{figure}

\end{document}